\documentclass[11pt, draftclsnofoot, onecolumn]{IEEEtran}
\IEEEoverridecommandlockouts
\usepackage{cite}
\usepackage{amsmath,amssymb,amsfonts}
\usepackage{algorithmic}
\usepackage{graphicx}
\usepackage{textcomp}
\usepackage{caption}
\usepackage{color}
\usepackage{xcolor}
\usepackage{subfigure}
\usepackage{booktabs}
\usepackage{setspace}
\usepackage{epstopdf}
\usepackage{multicol}
\usepackage{stfloats}

\usepackage[norelsize, linesnumbered, ruled, lined, boxed, commentsnumbered]{algorithm2e}
\SetKwRepeat{Do}{do}{while}%
\SetKw{Kwin}{ in }
\SetKw{KwTo}{ to }
\SetKw{KwGoto}{goto }

\newcommand{\EquRef}[1]{Eq. \ref{#1}}

\newcommand{\TableRef}[1]{Tab. \ref{#1}}
\newcommand{\FigRef}[1]{Fig. \ref{#1}}
\newcommand{\Athr}[1]{{#1} \emph{et al.}}
\newcommand{\AlgRef}[1]{Alg. \ref{#1}}

\def\BibTeX{{\rm B\kern-.05em{\sc i\kern-.025em b}\kern-.08em
    T\kern-.1667em\lower.7ex\hbox{E}\kern-.125emX}} 

\begin{document}

\title{Enhanced C-V2X Mode 4 to Optimize Age of Information and Reliability for IoV\\
\thanks{This work was supported in part by the National Natural Science Foundation of China (No. 61701197), in part by the open research fund of State Key Laboratory of Integrated Services Networks (No. ISN23-11), in part by the 111 Project (No. B23008). (Corresponding author: Qiong Wu)}
}

\author{\IEEEauthorblockN{Jiahou Chu$^{1,2}$, Qiong Wu$^{1,2}$, Qiang Fan$^{3}$ and Zhengquan Li$^{1}$}\\
	\IEEEauthorblockA{\small $^{1}$ School of Internet of Things Engineering, Jiangnan University, Wuxi 214122, China}\\
	\IEEEauthorblockA{\small $^{2}$ State Key Laboratory of Integrated Services Network (Xidian University), Xi'an 710071, China}\\
	\IEEEauthorblockA{\small $^3$ Qualcomm, San Jose CA 95110 USA, China}\\
	\IEEEauthorblockA{\small Email: qiongwu@jiangnan.edu.cn}
	
}
\vspace{-10cm}	


\maketitle

\begin{abstract}
Internet of vehicles (IoV) has emerged as a key technology to realize real-time vehicular application.
For IoV, vehicles adopt cellular vehicle-to-everything (C-V2X) standard to support direct communication among them.
C-V2X mode 4 controls resource allocation without the assistance of cellular network, hence it is widely used for IoV.
However, C-V2X mode 4 has two drawbacks.
First is that vehicles cannot communicate with each other for a period in some case which will cause an increase in age of information (AoI); second is that vehicles may select resource already occupied by others which will deteriorate the reliability.
To address the two drawbacks, we propose an enhanced C-V2X mode 4 to optimize AoI and reliability.
In addition, we consider the fact that for most vehicular applications, each vehicle periodically requires fresh information of vehicles within a certain distance and propose a new performance metric to evaluate the system AoI for IoV.
Furthermore, we construct a platform through integrating SUMO and NS3.
We demonstrate the superiority of the enhanced C-V2X mode 4 base on this simulation platform.

\end{abstract}

\begin{IEEEkeywords}
IoV, C-V2X mode 4, AoI, reliability
\end{IEEEkeywords}

\section{Introduction}

With the development of internet of things (IoT), IoV has emerged as a key technology to realize real-time vehicular application\cite{wu2016performance} such as autonomous driving and navigation \cite{E220093, 9944845, 7248891, s18124198, 7036784}.
For IoV, vehicles adopt C-V2X standard to support direct communication among them \cite{3gpp.36.213}.
3rd generation partnership (3GPP) Release 14 has specified two modes for C-V2X standard, i.e., mode 3 and mode 4, to implement centralized and decentralized resource allocation, respectively \cite{3gpp.36.885}.
C-V2X mode 3 achieves higher performance than mode 4 due to the centralized resource allocate for vehicles based on cellular network. However, vehicles may not always stay in the coverage of a base station, which restricts the use of C-V2X mode 3.
C-V2X mode 4 controls resource allocation without the assistance of base stations, hence it is used widely for IoV.

C-V2X mode 4 has two drawbacks. First, each vehicle, adopting C-V2X mode 4, will select a resource and reserve a number of resources at equal time intervals. Due to half-duplex, if the resources two vehicles reserved overlap in time domain, they cannot communicate with each other for a period of time until one of they triggers resource reselection. This will degrade the information freshness, which is measured by AoI, i.e., the time difference between the current time and the information generation time \cite{6195689}. The second drawback is that vehicles may fail to determine which resources has been reserverd by other vehicles due to that each vehicle is not aware of the remaining quantity of the resources reserved by the other vehicles, thus it may select the resources which are already reserved by other vehicles and result in transmission collisions. This would decrease packet delivery rate (PDR) and deteriorate the reliability. The two drawbacks may prevent vehicles supporting the real-time vehicular applications which require low AoI and high reliability.

Some works have focused on improving the reliability of C-V2X mode 4. In \cite{9759361}, \Athr{Segawa} considered vehicle mobility and proposed a new algorithm to improve reliability through dynamically adjusting the packet transmission interval in C-V2X mode 4. In \cite{9323042}, \Athr{Kang} proposed adaptive transmission power and message interval control (ATMOIC) method, which adjusts the transmission power in physical layer of C-V2X mode 4 to improve reliability. However, these works did not taken AoI into account. Currently, some works have studied AoI of C-V2X Mode 4. In \cite{parvini2023aoi}, \Athr{Parvini} proposed two algorithms to improve average AoI by selecting resource and adjusting transmission power in C-V2X mode 4. In \cite{9214855}, \Athr{peng} proposed a resource scheduling method based on C-V2X mode 4 to improve the average AoI and PDR. To the best of our knowledge, no work has considered the two drawbacks of C-V2X mode 4 and proposed a new resource allocation method to jointly improve the AoI and reliability for IoV, which motivated us to conduct this work.


In this paper, we considered the two drawbacks of C-V2X mode 4 and proposed an enhanced C-V2X mode 4 to optimize AoI and reliability for IoV\footnote{The source code can be found at https://github.com/qiongwu86/ns3\_sumo\_cv2x\_mode4.git}. The main contributions are listed as follows

\begin{itemize}
	\item[1)] Enhanced C-V2X mode 4 is proposed. We first propose a new resource reservation method to optimize AoI. Then, we design a new sidelink control information (SCI) format to indicate the number of reserved resources, thus the reliability can be improved.
	\item[2)] Considering for most vehicular applications, each vehicle always requests fresh information of other vehicles within a certain distance and uses the information periodically, we propose a new metric, named $AoIS$, to measure the system AoI.
	\item[3)] A simulation platform to realize enhanced C-V2X mode 4 and the vehicle mobility has been constructed. We integrate SUMO and NS3 to construct the new platform and demonstrate the superiority of the enhanced C-V2X mode 4 based on it.
\end{itemize}

The remainder of this paper is organized as follows. Section II introduces C-V2X mode 4. Section III presents the enhanced C-V2X mode 4. Section IV presents the evaluation metrics and simulation results. Section V draws the conclusions.

\section{C-V2X Mode4}

%


C-V2X mode 4 utilizes the long term evolution sidelink (LTE-SL) technology at the physical layer to support 10MHz or 20MHz channel. The channel is divided into subchannels. For each transmission, a vehicle employs a single subframe resource (SSR) to broadcast a transport block (TB) and sidelink control information, where a SSR occupies a subframe, i.e., 1ms, in the time domain and a subchannel in the frequency domain, the TB and SCI are the data and control information, respectively. A resource block (RB) is the smallest resource unit in LTE-SL. For a SSR, 2 RBs is adopted to transmit the SCI and $N$ RBs are adopted to transmit the TB, here $N$ is defined in LTE-SL \cite{3gpp.36.331}.

\begin{figure}[htbp]
	\centering
	\includegraphics[scale=0.9]{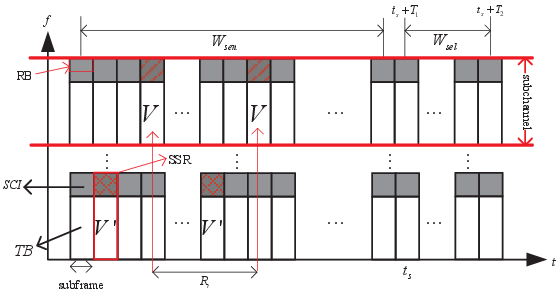}
	\caption{Process of SB-SPS}
	\label{SB-SPS}
\end{figure}

Each vehicle adopts the sensing based semi-persistent scheduling (SB-SPS) method to select a SSR and then reserves a certain numbers of SSRs to transmit consecutive packets, where the number is indicated by a counter, named reselection counter (RC). We consider each vehicle has a packet to transmit after every $R_t$ subframes. When a vehicle $V$ needs to transmit at subframe $t_s$ and RC is 0, it sets a random integer as RC within range $[a, b]$ and generates a value based on a uniform distribution within range $[0, 1]$. If the value is less than a predefined value $\beta$, it reuses the previous SSR for the following RC transmissions. Otherwise, the SB-SPS method is triggered to reselect a new SSR and then the selected SSR is reserved for the following RC transmissions.

Let the sensing window $W_{sen}$ be the subframes from $t_s-1000$ to $t_s-1$, the selection window $W_{sel}$ be the subframes from $t_s+T_{1}$ to $t_s+T_{2}$, where $T_{1}<4$ and $20 \leq T_{2} \leq 100$. Denote $S_{A}$ as the set of all SSRs in $W_{sel}$, $M_{total}$ as the total number of the SSRs in $S_{A}$ and $R_{x,y}$ as a SSR in $S_{A}$ which is located at subchannel $x$ and subframe $y$. Detailed process of SB-SPS is shown in \FigRef{SB-SPS} and we will introduce it as follows.

\begin{itemize}	
	\item[1)] Vehicle $V$ excludes the unusable SSRs from $S_{A}$. Specifically, vehicle $V$ first senses the received information in $W_{sen}$ and obtains the received signal strength indicators (RSSIs) for the corresponding SSRs. Then it selects the information whose SCI can be decoded and measures the reference signal received powers (RSRP) according to the corresponding SCI, where the set of the corresponding SSRs is denoted as $C_{sen}$. After that vehicle $V$ excludes the SSRs from $S_{A}$ if they satisfy either of the two following cases \cite{3gpp.36.213}.
	\begin{itemize}
		\item Vehicle $V$ needs to exclude the SSRs that may be occupied by other vehicles due to half-duplex.
		\item Vehicle $V$ needs to exclude the SSRs that will be occupied by other vehicles and suffer from interference of other vehicles, i.e., RSRP corresponding to the SSR is larger than the RSRP threshold which is predefined for SB-SPS\footnote{The first case means that vehicle cannot determine if the SSR has been occupied because they cannot receive information due to half-duplex and the second case means that the transmission will be severely affected if the selected SSR meets with this condition.}.
	\end{itemize}
	After excluding the aforementioned SSRs from $S_{A}$, vehicle $V$ calculates the number of SSRs remaining in $S_{A}$. If the value is less than $0.2*M_{total}$, $P_{TH}$ is set as $P_{TH}+3dB$ and step 1 is executed again. Otherwise, the process moves to step 2.
	\item[2)] Vehicle $V$ calculates average received signal strength indicator (A-RSSI) which reflects the level of interference for each SSR in $S_{A}$. Specifically, the A-RSSI of $R_{x,y}$ is the average of the corresponding received signal strength indicator (RSSI) of SSRs which are in $W_{sen}$ and spaced at integer multiples of 100 subframes from $R_{x,y}$.
	\item[3)] Vehicle $V$ selects the SSR for transmission. Specifically, according to the ascending order of A-RSSI, vehicle $V$ sequentially adds the corresponding SSRs in $S_{A}$ to the set $S_{B}$ until the number of SSRs in $S_{B}$ is larger than or equal to $0.2M_{total}$. Then it selects a SSR randomly from $S_{B}$.
\end{itemize}

Then vehicle $V$ reserves the following RC SSRs for RC transmission. After each transmission RC is decreased by 1. When RC is decreased to 0, vehicle $V$ resets a random integer as RC from $[a, b]$ and make a decision on whether to reselect a new SSR.

\section{Enhanced C-V2X Mode 4}

In this section, we will introduce the enhanced C-V2X mode 4 to optimize AoI and reliability.

We first propose an enhanced SB-SPS (ESB-SPS) to select SSR. The pseudocode of ESB-SPS is shown in \AlgRef{pseudocode of ESB-SPS}.
Suppose that vehicle $V$ triggers ESB-SPS. The inputs of ESB-SPS are $W_{sen}$, $C_{sen}$, $S_A$, RC, and $P_{TH}$ while the output is the selected SSR.
Vehicle $V$ first sets $M_{total}$ as the number of SSRs in $S_A$ (line 1).
Then, vehicle $V$ iteratively exclude SSRs from $S_{A}$ that may be occupied by other vehicles.
Similar with SB-SPS, the termination condition for this loop is that the number of SSRs in $S_A$ reaches $0.2 \times M_{total}$.

Next we introduce one iterative of the process, where $V$ will check all the SSRs in $S_A$ whether it should be excluded (lines 3-19).
Specifically, when vehicle $V$ is checking SSR $R$ from $S_A$, it will calculate the set $C_V$, which includes the SSRs that will be reserved if $V$ selects $R$.
However, the method to reserve SSRs is different from that in standard C-V2X mode 4.
Note that two vehicles cannot communicate to each other when they use the SSRs located on the different subchannels and same subframe due to half-duplex.
For the standard C-V2X mode 4, they would adopt the reserved SSRs which are also located on the different subchannels and same subframe for transmission until the value of RC for one vehicle is decreased to 0, which causes communication failure for a long time and thus deteriorate the AoI.
In order to solve this problem, we propose a novel resource reservation method.
Specifically, if vehicle $V$ selects a SSR $R$ in $S_A$, it will calculate the RC SSRs, which would be reserved, by calling functions $CO^{[i]}$ on $R$, where $i \in [0, RC-1]$, and then store them to set $C_V$.
For function $CO^{[i]}$, it will map $R$ to another SSR,
\begin{equation}
\begin{aligned}
	&CO^{[i]}[(x,y,z)]=
	&((x+i*\frac{R_t}{10})mod(1024), (y+i*z)mod(10), z),
\end{aligned}
\end{equation}
$x$, $y$ and $z$ represent the frame number, subframe number and subchannel number of $R$, respectively. Note that 1 frame consists of 10 subframes, i.e., 10ms, 1 system frame number (SFN) cycle consists of 1024 frames. Also considering there are $SC$ subchannels, we get $x \in [0,1023], y \in [0,9], z \in [0,SC-1]$.

By employing one of the reserved SSRs for transmission, when two vehicles use SSRs located on the different subchannels and same subframe to transmit, their SSR for next transmission will not be in the same subframe. Thus the time for the communication failure due to half-duplex can be reduced and AoI can be improved.

Then, vehicle $V$ will enter the loops to check if $R$ should be excluded (lines 8-18).
Specifically, for each SSR $R'$ in $C_{sen}$, let $RC'$ and $RSRP'$ be the corresponding RC and RSRP of $R'$.
Vehicle $V$ can extracts $RSRP'$ from the SCI of $R'$.
Moreover, vehicle $V$ also needs to obtain $RC'$.
However, for standard C-V2X mode 4, vehicle $V$ is unable to get RC directly from $R'$.
According to 3GPP \cite{3gpp.36.213}, vehicle $V$ estimates RC with the predicted value $\lceil 100/R_t \rceil$, which is not an accurate value.
Hence the vehicle $V$ cannot determine the rest SSRs that has been reserved by the sender of $R'$, which will reduce the accuracy in determining whether $R$ should be excluded and increase the probability of transmission failure.


For the standard C-V2X, the format of SCI includes the priority and retransmission information, which occupies 8 bits.
However, considering retransmission may increase the load of IoV\footnote{The vehicles, which adopt retransmission mechanism, will transmit a packet twice\cite{3gpp.36.213}.}, which will reduce the reliability when the number of vehicles increase.
Therefore, we consider the vehicles have the same priority and do not use retransmission mechanism.
In addition, according to the protocol specification, the maximum of RC ranges within $[25, 75]$ when $R_t=20$.
In this case, we need at least $\left\lceil \log_{2}{75-25+1} \right\rceil = 6$ bits to transmit RC, hence SCI can accommodate RC by excluding priority and retransmission information.
We replace the priority and retransmission information with RC to propose a new SCI format and $RC'$ can be obtained directly from the SCI of $R'$, hence the number of transmission failure can be reduced which improve the reliability.
The proposed SCI format is shown in \TableRef{sci}.

\begin{table}[htbp]
	\centering
	\caption{Proposed SCI format}
	\label{sci}
	\resizebox{0.7\textwidth}{!}{
	\begin{tabular}{ccc}
		\toprule
		Index & Item & bits \\
		\midrule
		1 & Resource Reservation &  4 \\
		2 & Frequency Resource Location & $\log(SC(SC+1)/2)$ \\
		3 & MCS & 5 \\
		4 & Transmission Format & 1\\
		5 & Reserved & $14 - \log(SC(SC+1)/2)$ \\
		6 & RC & 8  \\
		\bottomrule
	\end{tabular}}
\end{table}

Then vehicle $V$ calculetes the SSRs that will be reserved by vehicle $V'$ according to $CO$ function and stores them in a set $C_{V'}$ (lines 11-13).
After that, if vehicle $V$ and $V'$ may reserve the same SSR for transmission, i.e., $C_{V} \bigcap C_{V'} \neq \emptyset$, and vehicle $V$ is significantly interfered by vehicle $V'$, i.e., $RSRP' > P_{TH}$, vehicle $V$ will exclude $R$ from $S_A$.
After processing $R$, the algorithm will repeat the same procedure to check the next SSR in $S_A$.
When all SSRs in $S_A$ have been checked, vehicle $V$ increases $P_{TH}$ by 3dB. If the number of SSRs in $S_A$ is less than $0.2 \times M_{total}$, the above process is repeated. Otherwise, vehicle $V$ calculates the A-RSSI for each SSR in $S_A$ and sorts them (lines 20-22).


Specifically, to calculate A-RSSI of $R$, vehicle $V$ should select the RSSI of SSRs which is able to be mapped to $R$ by $CO$ function, thus the function to calculate A-RSSI of $R$, located at $(x,y,z)$, is defined as

\begin{equation}\label{asl-rssi}
	\begin{aligned}
    &AveRSSI(R)=
	&\frac{1}{|C_{RSSI}(R)|}\sum_{r \in C_{RSSI}(R)}^{}[RSSI(r)],
	\end{aligned}
\end{equation}
where
\begin{equation}
	   \begin{aligned}
		C_{RSSI}(R)=\{(a,b,c) \in W_{sen}|
		CO^{[i]}[(a,b,c)]=(x,y,z),\exists i \in N\}.
	   \end{aligned}
\end{equation}

$|C_{RSSI}(R)|$ is the number of elements in set $C_{RSSI}(R)$ and $RSSI(r)$ is RSSI of SSR $r$.

Afterwards, vehicle $V$ moves the SSRs with the lowest A-RSSI from $S_A$ to $S_B$ and selects a SSR from $S_B$ randomly as $R_{new}$ (lines 23-26). Then ESB-SPS is terminated and vehicle $V$ reserves RC SSRs based on function $CO$ for transmission.

\begin{algorithm}
    \setstretch{0.75}
    \LinesNumbered
    \caption{pseudocode of ESB-SPS}
    \label{pseudocode of ESB-SPS}
	\small
    \KwIn{
		$W_{sen}$, $C_{Sen}$, $S_{A}$, RC, $P_{TH}$
    }
    \KwOut{SSR $R_{new}$}
    $M_{total} = |S_{A}|$;\\
    \Do{$S_{A} < 0.2 \times M_{total}$}
    {
        \For{$R \Kwin S_{A}$}
        {
            $C_{V} = \emptyset$;\\
            \For
            {
                $i \Kwin [0, RC-1]$
            }
            {
                append $CO^{[i]}(R) \KwTo C_{V}$
            }
            \For{$R' \Kwin C_{sen}$}
            {
				Get $RC'$ and $RSRP'$ of $R'$;\\
                $C_{V'} = \emptyset$;\\
                \For{
                    $i \Kwin [0,RC'-1]$
                }
                {
                    appand $CO_{R'}^{[i]}$ \KwTo $C_{V'}$;\\
                }
                \If{$C_{V} \bigcap C_{V'} \neq \emptyset$ and $RSRP' > P_{TH}$}
                {
                    exclude $R$ from $S_{A}$;\\
                    \KwGoto \ref{OneSSR}
                }
            }\label{OneSSR}
        }
        $P_{TH} \gets P_{TH} + 3dB$;\\
    }
    Calculate $A-RSSI$ of each SSR in $S_{A}$;\\
    \Do{$S_{B} < 0.2 \times M_{total}$}
    {
        Move $R$, which has the smallest $A-RSSI$, to $S_{B}$
    }
    $R_{new} \gets $ random select in $S_{B}$;\\
    return $R_{new}$;
    \label{Over}
\end{algorithm}

\section{Simulation Results}

In this section, we first introduce the simulation scenario and relevant evaluation metrics we proposed. Then, we present and analyze the simulation results.

\subsection{Simulation Scenario and Evaluation Metrics}

\begin{figure}[htbp]
	\centering
	\rotatebox{0}{\includegraphics[width=0.53\textwidth]{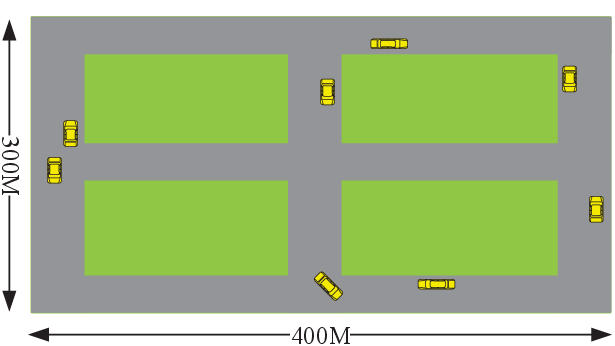}}
	\caption{Simulation scenario}
	\label{system model}
\end{figure}

We construct a simulation platform by combining NS3 and SUMO, where NS3 interacts with SUMO through the TraCi library to obtain real-time positions of vehicles.
The mobility model is the \emph{Manhattan Mobility Model} in SUMO, where vehicles drive in the Manhattan street grid, which is a typical urban scenario.
Vehicles are controlled by the car following model named Krauss in SUMO when they are driving along the road. Each vehicle has the options to turn left, turn right and go straight at an intersection with probabilities 0.25, 0.25, and 0.5, respectively.
Each vehicle adopts the enhanced C-V2X mode 4 to broadcast packets.
The communication range of each vehicle is large enough.
The enhanced C-V2X mode 4 is designed based on the C-V2X mode 4 in NS3 proposed by Eckermann \cite{Eckermann2019performance}.
In addition, because C-V2X is a service designed to provide high reliability and timeliness in a certain distance, we consider that each vehicle only needs the information which satisfies the following two conditions. (1) The information is fresh enough, i.e., the AoI of the information is less than a AoI threshold $AoI_{th}$; (2) the information is obtained from the vehicles within a distance $d$. Each vehicle checks the two conditions every 100 ms. The following two metrics are used to evaluate the reliability and timeliness of the system, respectively. The simulation scenario is shown in \FigRef{system model} and the simulation parameters are listed in \TableRef{sim param}.

\vspace{-0.4cm}	

\begin{itemize}
	\item \emph{PDR}

	\hspace{1em} PDR is calculated as the proportion of the successfully transmitted packets when the distance between the sender and receiver pairs less than $d$, which is an important metric to measure network reliability.

	\item \emph{System AoI}

	\hspace{1em}
	We define the proportion of AoI that is higher than $AoI_{th}$ to reflect the system AoI, i.e.,
	\begin{equation}	
	\begin{aligned}
		AoIS(AoI_{th},d)=\frac{\text{times of exceeding $AoI_{th}$}}{\text{times of checking AoI}} \times 100\%.
	\end{aligned}\label{calPDR}
	\end{equation}

	\hspace{1em}
	Note that each vehicle only checks the AoI of the information received from vehicles within a distance $d$ every 100 ms. In Eq. (4), denominator is the times that all vehicles check AoI of the information and the numerator is the times that AoI is higher than $AoI_{th}$.
\end{itemize}

\subsection{Simulation Results}
%


\begin{table}[htbp]
  	\centering
	\caption{simulation parameter}
	\label{sim param}
	\resizebox{0.3\textwidth}{!}{
	\begin{tabular}{cl}
		\toprule
		\textbf{\MakeUppercase{GENERAL PARAMETER}} & \space \\
		\midrule
		channel model & WINNER+B1\\
		NS3 version & 3.28\\
		SUMO version & 1.18.0\\
		simulation time & 30 s\\
		\toprule
		\textbf{\MakeUppercase{V2X parameters}} & \space \\
		\midrule
		message size & 190 bytes\\
		transmission power & 23dBm\\
		$R_t$ & 20\\
		$T_{1},T_{2}$ & 4, 20\\
		$\beta$ & 0\\
		$SC$ & 3\\
		channel bandwidth & 10MHz\\
		subframe bitmap & 0xFFFFF\\
		subchannel scheme & adjacent\\
		\bottomrule
	\end{tabular}
	}
\end{table}

\begin{figure}
	\centering
	\includegraphics[scale=0.5]{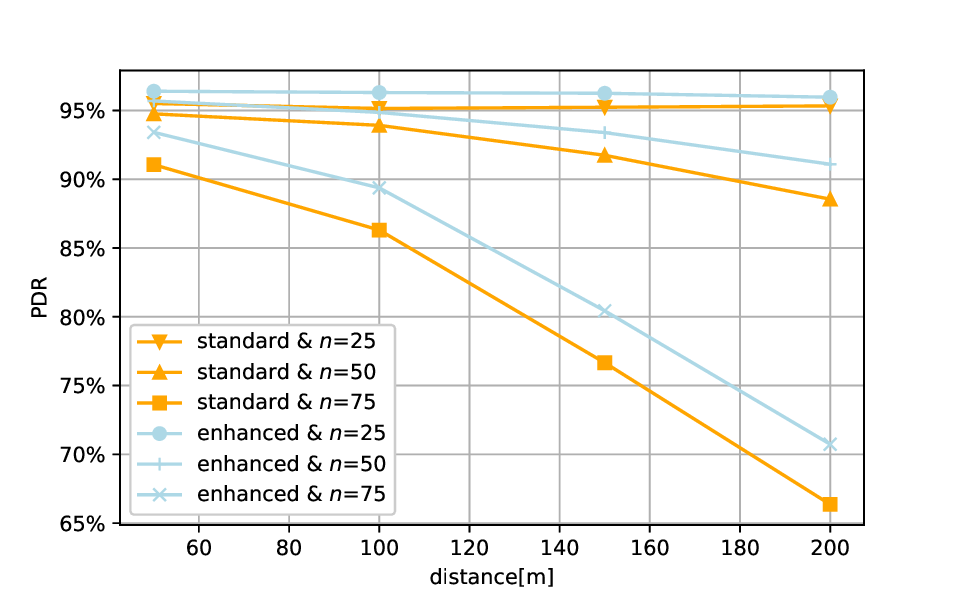}
	\caption{PDR of standard and enhanced C-V2X mode 4 for 25, 50 and 75 vehicles numbers.}
	\label{PDR}
\end{figure}

\FigRef{PDR} shows PDR under different distance $d$ when the numbers of vehicles, denoted as $n$, are 25, 50 and 75, respectively.
We can see that the enhanced C-V2X mode 4 outperforms the standard C-V2X mode 4.
It is because that RC is transmitted in the SCI for enhanced C-V2X mode 4, and it will be more accurate to exclude SSRs that may be reserved by other vehicles.
In addition, the PDR of both C-V2X mode 4 decreases as $d$ increases. It is because that the transmission power is fixed and the signal to noise ratio (SNR) will decrease as $d$ increase, which will lead to a decrease in PDR.
In addition, we can also see that PDR increases as the number of vehicles increases for both standard and enhanced C-V2X mode 4.
It is because that the number of available SSRs is fixed, the system would become congested as the number of vehicles increases and thus reduce PDR.

\begin{figure*}[thbp]
	\centering
	\includegraphics[width=1.0\textwidth]{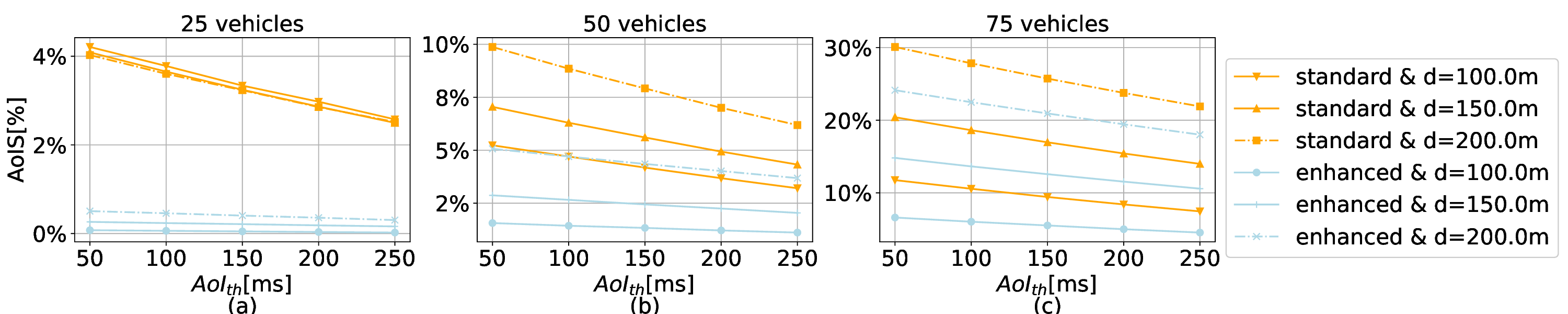}
	\caption{System AoI of standard and enhanced C-V2X mode 4 for 25, 50 and 75 vehicles.}
	\label{aoioverrate}
\end{figure*}

\FigRef{aoioverrate} compares the system AoI, i.e., $AoIS$, of enhanced C-V2X and standard C-V2X under different $AoI_{th}$ and $d$ when the numbers of vehicles are 25, 50, and 75, respectively.
Firstly, we can see the $AoIS$ of enhanced C-V2X mode 4 is always less than that of standard C-V2X mode 4.
It is because that the resource reservation method used in enhanced C-V2X mode 4 will alleviate the drawback of C-V2X mode 4 caused by half-duplex, and reduces $AoIS$.
In addition, we can also see that $AoIS$ of both enhanced and standard C-V2X mode 4 increases as $d$ and number of vehicles increase. This is because that according to \FigRef{PDR}, PDR decreases as $d$ and number of vehicles increases. In this case information cannot be updated timely, thus ultimately increasing $AoIS$. Moreover, we can see that $AoIS$ of both enhanced and standard C-V2X mode 4 decreases when $AoI_{th}$ increases.  This is because that the numerator of \EquRef{calPDR} decreases as $AoI_{th}$ increases, which leads to a decrease of $AoIS$.

\section{Conclusions}
In this paper, we proposed enhanced C-V2X mode 4 to optimize the AoI and reliability for IoV. We first proposed an enhanced C-V2X mode 4 to address the two drawbacks of C-V2X.
Then we proposed a new performance metric to measure the system AoI for IoV.
After that we constructed a platform by integrating SUMO and NS3 to realize enhanced C-V2X mode 4 and mobility of vehicles simultaneously.
Finally, we demonstrated the superiority of the enhanced C-V2X mode 4 base on this simulation platform.



\bibliographystyle{IEEEtran}
\bibliography{RefsFile}

\end{document}